\documentclass[11pt]{article}
\usepackage{amssymb,amsmath,amsfonts}
\usepackage{graphicx}
\usepackage{graphics}
\usepackage{eepic,epsfig}

\begin{document}

\title{Electromagnetic two-point functions and Casimir densities \\
for a conducting plate in de Sitter spacetime}
\author{A. A. Saharian\thanks{%
E-mail: saharian@ysu.am}, A. S. Kotanjyan, H. A. Nersisyan \\
\\
\textit{Department of Physics, Yerevan State University,}\\
\textit{1 Alex Manoogian Street, 0025 Yerevan, Armenia}}
\maketitle

\begin{abstract}
We evaluate the two-point function for the electromagnetic field tensor in $%
(D+1)$-dimensional de Sitter spacetime assuming that the field is prepared
in Bunch-Davies vacuum state. This two-point function is used for the
investigation of the vacuum expectation values (VEVs) of the field squared
and the energy-momentum tensor in the presence of a conducting plate. The
VEVs are decomposed into the boundary-free and plate-induced parts. For the
latter, closed form analytical expressions are given in terms of the
hypergeometric function. For $3\leqslant D\leqslant 8$ the plate-induced
part in the VEV of the electric field squared is positive everywhere,
whereas for $D\geqslant 9$ it is positive near the plate and negative at
large distances. The VEV of the energy-momentum tensor, in addition to the
diagonal components, contains an off-diagonal component which corresponds to
the energy flux along the direction normal to the plate. Simple asymptotic
expressions are provided at small and large distances from the plate
compared with the de Sitter curvature scale. For $D\geqslant 4$, all the
diagonal components of the plate-induced vacuum energy-momentum tensor are
negative and the energy flux is directed from the plate.
\end{abstract}

\bigskip


\bigskip

\section{Introduction}

It is well-known that the properties of vacuum state for a quantum field
crucially depend on the geometry of background spacetime. Closed expressions
for physical characteristics of the vacuum, such as expectation values of
various bilinear products of the field operator, can be found for highly
symmetric background geometries only. On one hand these analytic results are
interesting on their own and on the other, they help understanding the
influence of the gravitational field on the quantum vacuum for more
complicated geometries. From this perspective, de Sitter (dS) spacetime is
among the most interesting backgrounds. There are several reasons for that.
dS spacetime is the maximally symmetric solution of Einstein's equations
with a positive cosmological constant and, owing to this symmetry, numerous
physical problems are exactly solvable on this background. In accordance
with the inflationary cosmology scenario \cite{Lind90}, in the early stages
of the cosmological expansion our universe passed through a phase in which
its local geometry closely resembles that of dS spacetime. During an
inflationary epoch, quantum fluctuations in the inflaton field introduce
inhomogeneities which play a central role in the generation of cosmic
structures from inflation. More recently, astronomical observations of high
redshift supernovae, galaxy clusters and cosmic microwave background \cite%
{Ries07} indicated that at the present epoch the universe is accelerating
and can be approximated by a world with a positive cosmological constant. If
the universe were to accelerate indefinitely, the standard cosmology would
lead to an asymptotic dS universe. It is therefore important to investigate
physical effects in dS spacetime for understanding both the early universe
and its future.

The interaction of a fluctuating quantum field with the background
gravitational field leads to vacuum polarization. The boundary conditions
imposed on the field operator give rise to another type of vacuum
polarization. These conditions can arise because of the presence of
boundaries having different physical natures, like macroscopic bodies in
QED, extended topological defects, horizons and branes in higher-dimensional
models. They modify the zero-point modes of a quantized field and, as a
result, forces arise acting on constraining boundaries. This is the familiar
Casimir effect (for reviews see \cite{Eliz94}). The particular features of
the Casimir forces depend on the nature of a quantum field, the type of
spacetime manifold, the boundary geometry, and the specific boundary
conditions imposed on the field.

An interesting topic in the investigation of the Casimir effect is its
explicit dependence on the geometry of the background spacetime. In the
present paper, an exactly solvable problem with both types of polarization
of the electromagnetic vacuum will be considered. We evaluate the two-point
function for the field tensor and the vacuum expectation values (VEVs) of
the field squared and the energy-momentum tensor induced by a conducting
plate in $(D+1)$-dimensional dS spacetime. The corresponding problem for a
massive scalar field with general curvature coupling parameter has been
recently considered in \cite{Saha09} and \cite{Milt12} for flat and
spherical boundaries respectively (see also \cite{Seta01,Burd11} for special
cases of conformally and minimally coupled massless fields). It has been
shown that the curvature of the background spacetime decisively influences
the behavior of VEVs at distances larger than the curvature scale of dS
spacetime. In another class of models with boundary conditions, the latter
arise because of the nontrivial topology of the space. The periodicity
conditions imposed on the field operator along compact dimensions lead to
the topological Casimir effect. This effect for scalar and fermionic fields,
induced by toroidal compactification of spatial dimensions in dS spacetime
has been investigated in \cite{Saha08}.

The outline of the paper is as follows. In the next section we present a
complete set of mode functions for the electromagnetic field defining the
Bunch-Davies vacuum state in $(D+1)$-dimensional boundary-free dS spactime.
Then, these mode function are used for the evaluation of the two-point
function of the electromagnetic field tensor. In Section \ref{sec:Plate} we
consider the geometry with a conducting plate. The corresponding mode
functions for the vector potential are presented and they are used for the
evaluation of the two-point functions. The latter are expressed in terms of
the boundary-free two-point functions. Then, we evaluate the parts in the
VEVs of the electromagnetic field squared and the energy-momentum tensor
induced by the conducting plate. Closed expressions for these VEVs are
derived and their asymptotic behavior is investigated near the plate and at
large distances compared with the dS curvatures scale. The main results are
summarized in Section \ref{sec:Conc}.

\section{Electromagnetic modes and two-point functions}

\label{sec:Modes}

In this section we present a complete set of mode functions for the
electromagnetic field in dS spacetime and then the two-point functions for
the field tensor are evaluated. The two-point functions $\langle A_{\mu
}(x)A_{\nu }(x^{\prime })\rangle $ for both massive and massless vector
fields in dS spacetime are obtained in \cite{Alle86} by using the arguments
based on the maximal symmetry. In \cite{Gaze00}, the Wightman two-point
functions for massive and massless vector fields are investigated in
4-dimensional dS spacetime by making use of the construction based on
analyticity properties offered by the complexified pseudo-Riemannian
manifold in which the dS manifold is embedded. Infrared pathologies in the
behavior of the photon two-point functions in dS spacetime have been
recently discussed in \cite{Yous11} and it has been shown that these are
purely gauge artifacts.

\subsection{Mode functions}

We consider the electromagnetic field in the background of $(D+1)$%
-dimensional dS spacetime, described in the inflationary coordinates:%
\begin{equation}
ds^{2}=dt^{2}-e^{2t/\alpha }\sum_{l=1}^{D}(dz^{l})^{2},  \label{ds2}
\end{equation}%
where the parameter $\alpha $ is related to the positive cosmological
constant $\Lambda $ by the formula $\alpha ^{2}=D(D-1)/(2\Lambda )$. Below,
in addition to the comoving synchronous time coordinate $t$ we use the
conformal time $\tau $, defined as $\tau =-\alpha e^{-t/\alpha }$, $-\infty
<\tau <0$. In terms of this coordinate the metric tensor takes a conformally
flat form: $g_{ik}=(\alpha /\tau )^{2}\mathrm{diag}(1,-1,...,-1)$. For the
action integral of the electromagnetic field one has%
\begin{equation}
S=-\frac{1}{16\pi }\int d^{D+1}x\,\sqrt{|g|}F_{\mu \nu }(x)F^{\mu \nu }(x),
\label{action}
\end{equation}%
where $F_{\mu \nu }$ is the electromagnetic field tensor: $F_{\mu \nu
}=\partial _{\mu }A_{\nu }-\partial _{\nu }A_{\mu }$. Imposing the gauge
conditions $A_{0}=0$,$\;\nabla _{\mu }A^{\mu }=0$, we expand the vector
potential as a Fourier integral%
\begin{equation}
A_{l}(x)=\int d\mathbf{k}\,A_{l}(\tau ,\mathbf{k})e^{i\mathbf{k}\cdot
\mathbf{z}},\;A_{l}(\tau ,-\mathbf{k})=A_{l}^{\ast }(\tau ,\mathbf{k}),
\label{potfurie}
\end{equation}%
with, $x=(\tau ,\mathbf{z})$, $\mathbf{z}=(z^{1},\ldots ,z^{D})$, $\mathbf{k}%
=(k_{1},\ldots ,k_{D})$ and $\mathbf{k}\cdot \mathbf{z}%
=\sum_{l=1}^{D}k_{l}z^{l}$. From the gauge condition, for the Fourier
components of the vector potential one has $\sum_{l=1}^{D}k_{l}A_{l}(\tau ,%
\mathbf{k})=0$.

In terms of the Fourier components, the action integral (\ref{action}) is
written in the form
\begin{equation}
S=\frac{(2\pi )^{D-1}}{4}\sum_{l=1}^{D}\int d\mathbf{k}\,\int d\tau \,\left(
\alpha /\eta \right) ^{D-3}\left[ \partial _{\tau }A_{l}(\tau ,\mathbf{k}%
)\partial _{\tau }A_{l}^{\ast }(\tau ,\mathbf{k})-k^{2}A_{l}(\tau ,\mathbf{k}%
)A_{l}^{\ast }(\tau ,\mathbf{k})\right] ,  \label{action2}
\end{equation}%
where $\eta =|\tau |$. The variational principle applied to (\ref{action2})
leads to the equation:
\begin{equation}
\partial _{\tau }^{2}A_{l}(\tau ,\mathbf{k})-\frac{D-3}{\tau }\partial
_{\tau }A_{l}(\tau ,\mathbf{k})+k^{2}A_{l}(\tau ,\mathbf{k})=0.  \label{AlEq}
\end{equation}%
The general solution of this equation is a linear combination of the
functions $\eta ^{D/2-1}H_{D/2-1}^{(1)}(k\tau )$ and $\eta
^{D/2-1}H_{D/2-1}^{(2)}(k\tau )$, with $H_{\nu }^{(1,2)}(x)$ being the
Hankel functions. Different choices of the coefficients in the linear
combination correspond to different vacuum states. Here we assume that the
field is prepared in dS invariant Bunch-Davies vacuum state which is the
only dS invariant vacuum state with the same short-distance structure as the
Minkowksi vacuum.

For Bunch-Davies vacuum state $A_{l}(\tau ,\mathbf{k})\propto \eta
^{D/2-1}H_{D/2-1}^{(2)}(k\tau )$. By taking into account that $H_{\nu
}^{(2)}(k\tau )=-e^{\nu \pi i}H_{\nu }^{(1)}(k\eta )$, the complete set of
mode functions defining this state are given by
\begin{equation}
A_{(\sigma \mathbf{k})l}(x)=C\epsilon _{(\sigma )l}\eta
^{D/2-1}H_{D/2-1}^{(1)}(k\eta )e^{i\mathbf{k}\cdot \mathbf{z}},\text{ }%
l=1,\ldots ,D,  \label{Amodes}
\end{equation}%
where $\sigma =1,\ldots ,D-1$ correspond to different polarizations. The
polarization vectors obey the relations%
\begin{equation}
\sum_{l=1}^{D}\epsilon _{(\sigma )l}k_{l}=0,\;\sum_{l=1}^{D}\epsilon
_{(\sigma )l}\epsilon _{(\sigma ^{\prime })l}=\delta _{\sigma \sigma
^{\prime }},  \label{Eps1}
\end{equation}%
and%
\begin{equation}
\sum_{\sigma =1}^{D-1}\epsilon _{(\sigma )l}\epsilon _{(\sigma )m}=\delta
_{lm}-k_{l}k_{m}/k^{2}.  \label{Eps2}
\end{equation}%
The normalization factor $C$ is determined from the condition
\begin{equation}
\int d\mathbf{z}\,\sum_{l=1}^{D}[A_{(\sigma \mathbf{k})l}^{\ast }(x)\partial
_{\tau }\ A_{(\sigma ^{\prime }\mathbf{k}^{\prime })l}(x)-A_{(\sigma
^{\prime }\mathbf{k}^{\prime })_{l}}(x)\partial _{\tau }\ A_{(\sigma \mathbf{%
k})l}^{\ast }(x)]=-i\frac{4\pi \delta _{\sigma \sigma ^{\prime }}}{(\alpha
/\eta )^{D-3}}\delta (\mathbf{k}-\mathbf{k}^{\prime }).  \label{NormCond}
\end{equation}%
By using the expression (\ref{Amodes}), one finds%
\begin{equation}
\left\vert C\right\vert ^{2}=\frac{\alpha ^{3-D}}{4(2\pi )^{D-2}}.
\label{C2dS}
\end{equation}%
For $D=3$, by taking into account that $H_{1/2}^{(1)}(z)=-i\sqrt{2/(\pi x)}%
e^{iz}$, we see that the mode functions coincide with the corresponding
functions in Minkowski spacetime (the $D=3$ mode functions have also been
considered in \cite{Cota10}). This is a consequence of the conformal
invariance of the electromagnetic field in $D=3$.

\subsection{Two-point functions}

\label{sec:2pFunc}

Having a complete set of normalized mode functions for the vector potential
we can evaluate the two-point function for the electromagnetic field by
using the mode-sum formula (here and below the Latin indices for tensors run
over $1,2,\ldots ,D$)%
\begin{equation}
\langle A_{l}(x)A_{m}(x^{\prime })\rangle _{0}=\sum_{\sigma =1}^{D-1}\int d%
\mathbf{k}\,A_{(\sigma \mathbf{k})l}(x)A_{(\sigma \mathbf{k})m}^{\ast
}(x^{\prime }),  \label{AlAmMode}
\end{equation}%
where $\langle \cdots \rangle _{0}$ stands for the VEV in the boundary-free
dS spacetime. By taking into account the expression (\ref{Amodes}), one gets
the integral representation
\begin{equation}
\langle A_{l}(x)A_{m}(x^{\prime })\rangle _{0}=\frac{\alpha ^{3-D}(\eta \eta
^{\prime })^{D/2-1}}{4(2\pi )^{D-2}}\int d\mathbf{k}\,\left( \delta _{lm}-%
\frac{k_{l}k_{m}}{k^{2}}\right) H_{D/2-1}^{(2)}(k\eta )H_{D/2-1}^{(1)}(k\eta
^{\prime })e^{i\mathbf{k}\cdot \Delta \mathbf{z}},  \label{AlAm}
\end{equation}%
with $\Delta \mathbf{z=z}-\mathbf{z}^{\prime }$.

First let us consider the part with $\delta _{lm}$. By writing the product
of the Hankel functions in terms of the Macdonald function, after the
integration over the angular part of $\mathbf{k}$, one finds:%
\begin{eqnarray}
\mathcal{I} &=&\int d\mathbf{k\,}H_{D/2-1}^{(2)}(k\eta
)H_{D/2-1}^{(1)}(k\eta ^{\prime })e^{i\mathbf{k}\cdot \Delta \mathbf{z}}
\notag \\
&=&\frac{4(2\pi )^{D/2}}{\pi ^{2}|\Delta \mathbf{z}|^{D/2-1}}%
\int_{0}^{\infty }dk\,k^{D/2}K_{D/2-1}(ik\eta )K_{D/2-1}(-ik\eta ^{\prime
})J_{D/2-1}(k|\Delta \mathbf{z}|),  \label{Hint}
\end{eqnarray}%
with $J_{\nu }(x)$ been the Bessel function. The last integral is evaluated
by using the formula from \cite{Prud86} and we get%
\begin{equation}
\mathcal{I}=4(2\pi )^{(D-3)/2}\frac{\Gamma (D-1)}{(\eta \eta ^{\prime
})^{D/2}}\frac{P_{(D-3)/2}^{(1-D)/2}(-Z)}{\left( Z^{2}-1\right) ^{(D-1)/4}},
\label{Hint2}
\end{equation}%
where $P_{\beta }^{\mu }(x)$ is the associated Legendre function,
\begin{equation}
Z=1+\frac{\left( \Delta \eta \right) ^{2}-|\Delta \mathbf{z}|^{2}}{2\eta
\eta ^{\prime }},  \label{Z}
\end{equation}%
and $\Delta \eta =\eta -\eta ^{\prime }$. The quantity $Z$ is invariant
under the action of the isometry group of dS spacetime. One has $Z>1$ and $%
Z<1$ for timelike and spacelike related points $x$ and $x^{\prime }$,
respectively. An alternative form for the integral (\ref{Hint2}) is obtained
by using the relation between the associated Legendre function and the
hypergeometric function \cite{Abra72}:
\begin{equation}
\mathcal{I}=\frac{2\pi ^{(D-3)/2}\Gamma (D-1)}{\Gamma \left( (D+1)/2\right)
(\eta \eta ^{\prime })^{D/2}}F\left( D-1,1;\frac{D+1}{2};z\right) ,
\label{Hint3}
\end{equation}%
with the notation%
\begin{equation}
z=\frac{Z-1}{2}=1+\frac{\left( \Delta \eta \right) ^{2}-|\Delta \mathbf{z}%
|^{2}}{4\eta \eta ^{\prime }}.  \label{z}
\end{equation}%
Hence, the two-point function for the vector potential is presented in the
form%
\begin{eqnarray}
\langle A_{l}(x)A_{m}(x^{\prime })\rangle _{0} &=&\frac{\delta _{lm}\alpha
^{3-D}\Gamma (D-1)}{(4\pi )^{(D-1)/2}\Gamma \left( (D+1)/2\right) }F\left(
D-1,1;\frac{D+1}{2};z\right)  \notag \\
&&-\frac{\alpha ^{3-D}(\eta \eta ^{\prime })^{D/2-1}}{4(2\pi )^{D-2}}\int d%
\mathbf{k}\,\frac{k_{l}k_{m}}{k^{2}}H_{D/2-1}^{(2)}(k\eta
)H_{D/2-1}^{(1)}(k\eta ^{\prime })e^{i\mathbf{k}\cdot \Delta \mathbf{z}}.
\label{AlAmFin}
\end{eqnarray}%
One has no closed form for the second term in the right-hand side of (\ref%
{AlAmFin}). However, this term does not contribute to the two-point
functions for the electromagnetic field tensor and it will not be needed in
the further consideration.

By using (\ref{AlAmFin}), we find the following expressions for the
two-point functions of the electromagnetic field tensor:
\begin{eqnarray}
\langle F_{0l}(x)F_{0m}(x^{\prime })\rangle _{0} &=&\frac{(\eta \eta
^{\prime })^{-2}}{2B_{D}\alpha ^{D-3}}\left[ \left( \delta _{lp}\delta
_{mq}-\delta _{lm}\delta _{pq}\right) \frac{\Delta z^{p}\Delta z^{q}}{2\eta
\eta ^{\prime }}\partial _{z}+\left( D-1\right) \delta _{lm}\right] G_{D}(z),
\notag \\
\langle F_{pl}(x)F_{0m}(x^{\prime })\rangle _{0} &=&\frac{(\eta \eta
^{\prime })^{-2}}{B_{D}\alpha ^{D-3}}\delta _{\lbrack pm}\delta _{l]q}\frac{%
\Delta z^{q}}{\eta ^{\prime }}\left[ 2+\left( z-\frac{\eta +\eta ^{\prime }}{%
2\eta }\right) \partial _{z}\right] F_{D}(z),  \label{F2p} \\
\langle F_{pl}(x)F_{qm}(x^{\prime })\rangle _{0} &=&\frac{(\eta \eta
^{\prime })^{-2}}{B_{D}\alpha ^{D-3}}\left( \delta _{\lbrack pr}\delta
_{l][q}\delta _{m]s}\frac{\Delta z^{r}\Delta z^{s}}{\eta \eta ^{\prime }}%
\partial _{z}+2\delta _{\lbrack pq}\delta _{l]m}\right) F_{D}(z),  \notag
\end{eqnarray}%
where%
\begin{equation}
B_{D}=(4\pi )^{(D-1)/2}\Gamma \left( (D+3)/2\right) ,  \label{BD}
\end{equation}%
and the square brackets enclosing the indices mean the antisymmetrization
over these indices: $a_{\cdots \lbrack i_{j}\cdots i_{k}]\cdots }=(a_{\cdots
i_{j}\cdots i_{k}\cdots }-a_{\cdots i_{k}\cdots i_{j}\cdots })/2$. In (\ref%
{F2p}) we have introduced the functions%
\begin{eqnarray}
F_{D}(z) &=&\Gamma (D)F\left( D,2;\frac{D+3}{2};z\right) ,  \notag \\
G_{D}(z) &=&2\Gamma (D-1)F\left( D-1,3;\frac{D+3}{2};z\right) .  \label{GD}
\end{eqnarray}%
For odd values of $D$ one has simple expressions:%
\begin{eqnarray}
F_{3}(z) &=&G_{3}(z)=2(z-1)^{-2},  \notag \\
F_{5}(z) &=&12(z-2)(z-1)^{-3},\;G_{5}(z)=-12(z-1)^{-3},  \label{F5} \\
F_{7}(z) &=&144\frac{2z^{2}-6z+5}{(z-1)^{4}},\;G_{7}(z)=-48\frac{2z-5}{%
(z-1)^{4}}.  \notag
\end{eqnarray}%
The expression for the two-point function $\langle F_{0m}(x)F_{pl}(x^{\prime
})\rangle $ is obtained from the expression (\ref{F2p}) for $\langle
F_{pl}(x)F_{0m}(x^{\prime })\rangle $ by changing the sign and by the
interchange $\eta \rightleftarrows \eta ^{\prime }$.

\section{Two-point functions and Casimir densities in the geometry with a
conducting plate}

\label{sec:Plate}

\subsection{Two-point functions}

As an application of the formulas for the two-point functions given above
here we consider the change in the properties of the electromagnetic vacuum
induced by the presence of a perfectly conducting plate placed at $z^{D}=0$
(for the electromagnetic Casimir effect in higher-dimensional spacetimes
see, for instance, \cite{Ambj83,Alne07}). We consider the region $z^{D}>0$.
On the plate the field obeys the boundary condition \cite{Ambj83} $n^{\nu
_{1}}\,^{\ast }F_{\nu _{1}\cdots \nu _{D-1}}=0$, with the tensor $^{\ast
}F_{\nu _{1}\cdots \nu _{D-1}}$ dual to $F_{\mu \nu }$, and $n^{\mu }$ is
the normal to the plate. The corresponding mode-functions for the vector
potential are given by the expressions%
\begin{eqnarray}
A_{(\sigma \mathbf{k})l}(x) &=&iC_{b}\epsilon _{(\sigma )l}\eta
^{D/2-1}H_{D/2-1}^{(2)}(k\tau )\sin (k_{D}z^{D})e^{i\mathbf{k}_{\parallel
}\cdot \mathbf{z}_{\parallel }},  \notag \\
A_{(\sigma \mathbf{k})D}(x) &=&C_{b}\epsilon _{(\sigma )D}\eta
^{D/2-1}H_{D/2-1}^{(2)}(k\tau )\cos (k_{D}z^{D})e^{i\mathbf{k}_{\parallel
}\cdot \mathbf{z}_{\parallel }},  \label{AmodesB}
\end{eqnarray}%
where $l=1,\ldots ,D-1$, $\mathbf{k}_{\parallel }=(k_{1},\ldots ,k_{D-1})$, $%
\mathbf{z}_{\parallel }=(z^{1},\ldots ,z^{D-1})$ and $k=\sqrt{k_{D}^{2}+%
\mathbf{k}_{\parallel }^{2}}$. The polarization vectors obey the same
relations (\ref{Eps1}) and (\ref{Eps2}). The normalization coefficient is
determined from the condition (\ref{NormCond}), with the difference that now
the integration goes over the region $z^{D}>0$. In this way, we can see that
$\left\vert C_{b}\right\vert ^{2}=4\left\vert C\right\vert ^{2}$, where $%
\left\vert C\right\vert ^{2}$ is given by the expression (\ref{C2dS}).

The two-point functions for the vector potential are evaluated by using the
mode-sum formula similar to (\ref{AlAmMode}). Substituting the mode
functions, the two-point functions are presented in the decomposed form%
\begin{equation}
\langle A_{l}(x)A_{m}(x^{\prime })\rangle =\langle A_{l}(x)A_{m}(x^{\prime
})\rangle _{0}+\langle A_{l}(x)A_{m}(x^{\prime })\rangle _{\mathrm{b}},
\label{AlAmdec}
\end{equation}%
where the second term in the right-hand side is induced by the presence of
the conducting plate. For the latter one finds%
\begin{eqnarray}
\langle A_{l}(x)A_{m}(x^{\prime })\rangle _{\mathrm{b}} &=&-\langle
A_{l}(x)A_{m}(x_{-}^{\prime })\rangle _{0},  \notag \\
\langle A_{l}(x)A_{D}(x^{\prime })\rangle _{\mathrm{b}} &=&\langle
A_{l}(x)A_{D}(x_{-}^{\prime })\rangle _{0},  \label{AlAmb}
\end{eqnarray}%
where $l=1,\ldots ,D$, $m=1,\ldots ,D-1$, and $x_{-}^{\prime }$ is the image
of $x^{\prime }$ with respect to the plate: $x_{-}^{\prime }=(\tau ^{\prime
},z^{1\prime },\ldots ,z^{D-1\prime },-z^{D\prime })$. For the two-point
function of the electromagnetic field tensor in the region $z^{D}>0$ we get
a similar decomposition%
\begin{equation}
\langle F_{pl}(x)F_{qm}(x^{\prime })\rangle =\langle
F_{pl}(x)F_{qm}(x^{\prime })\rangle _{0}+\langle F_{pl}(x)F_{qm}(x^{\prime
})\rangle _{\mathrm{b}},  \label{FFdecomp}
\end{equation}%
with the plate-induced parts given by%
\begin{eqnarray}
\langle F_{pl}(x)F_{qm}(x^{\prime })\rangle _{\mathrm{b}} &=&-\langle
F_{pl}(x)F_{qm}(x_{-}^{\prime })\rangle _{0},  \notag \\
\langle F_{pl}(x)F_{Dm}(x^{\prime })\rangle _{\mathrm{b}} &=&\langle
F_{pl}(x)F_{Dm}(x_{-}^{\prime })\rangle _{0},  \label{FFb}
\end{eqnarray}%
with $p,l=0,1,\ldots ,D$, and $q,m=0,1,\ldots ,D-1$.

Here we have considered the perfectly conducting boundary condition. The
case of the infinitely permeable boundary condition, $n^{\mu }F_{\mu \nu }=0$%
, is treated in a similar way. This condition is imposed in the bag model
for hadrons.

\subsection{Casimir densities}

We consider a free field theory and the two-point functions given above
encode all the properties of the vacuum state. In particular, having these
functions we can evaluate the VEVs for various physical observables
characterizing the vacuum state. First let us consider the VEV of the
electric field squared. It can be obtained from the two-point functions
given above in the coincidence limit of the arguments%
\begin{equation}
\langle E^{2}\rangle =-g^{00}g^{lm}\lim_{x^{\prime }\rightarrow x}\langle
F_{0l}(x)F_{0m}(x^{\prime })\rangle .  \label{E2}
\end{equation}%
On the base of the decomposition (\ref{FFdecomp}) we can write a similar
decomposition for the electric field squared: $\langle E^{2}\rangle =\langle
E^{2}\rangle _{0}+\langle E^{2}\rangle _{\mathrm{b}}$. For points away from
the plate the divergences in the coincidence limit of the two-point
functions are contained in the boundary-free part $\langle E^{2}\rangle _{0}$
only and, hence, the renormalization is needed for this part only. Because
of the maximal symmetry of dS spacetime and the vacuum state, this part does
not depend on the spacetime point.

The part induced by the plate is directly evaluated using the term in the
two-point function, (\ref{FFb}). By taking into account the first expression
from (\ref{F2p}), one gets%
\begin{equation}
\langle E^{2}\rangle _{\mathrm{b}}=\frac{D-1}{2B_{D}\alpha ^{D+1}}\left[
2(1-y)\partial _{y}-D+2\right] G_{D}(y),  \label{E2b}
\end{equation}%
with the notation%
\begin{equation}
y=1-\left( z^{D}/\eta \right) ^{2}.  \label{y}
\end{equation}%
The plate-induced part (\ref{E2b}) depends on $z^{D}$ and $\eta $ in the
combination $z^{D}/\eta $. The latter is the proper distance from the plate
measured in units of $\alpha $. For $D=3$, from the general result (\ref{E2b}%
) one finds $\langle E^{2}\rangle _{\mathrm{b}}=3(\alpha z^{D}/\eta
)^{-4}/(4\pi )$. The latter is obtained from the corresponding result in
Minkowski spacetime by the standard conformal transformation.

Let us consider the behavior of the plate-induced part in the VEV of the
field squared in the asymptotic regions of the ratio $z^{D}/\eta $. At
proper distances from the plate much smaller than the dS curvature scale $%
\alpha $ one has $z^{D}/\eta \ll 1$. By using the asymptotic formula
\begin{equation}
G_{D}(y)\approx \Gamma \left( \frac{D+3}{2}\right) \Gamma \left( \frac{D+1}{2%
}\right) (z^{D}/\eta )^{-D-1},  \label{GDAs}
\end{equation}%
we get%
\begin{equation}
\langle E^{2}\rangle _{\mathrm{b}}\approx \frac{3\left( D-1\right) \Gamma
((D+1)/2)}{2(4\pi )^{(D-1)/2}(\alpha z^{D}/\eta )^{D+1}}.  \label{E2bAs}
\end{equation}%
As it is seen, the plate-induced part diverges on the boundary. This type of
divergences are well-known in quantum field theory with boundaries. For
points near the plate the boundary-induced part (\ref{E2b}) dominates in the
VEV of the electric field squared.

At distances from the plate much larger than the dS curvature scale, $%
z^{D}/\eta \gg 1$, one has $y\ll -1$. In this case, we use the asymptotic
expressions%
\begin{eqnarray}
G_{D}(y) &\approx &\frac{2^{D-4}}{\sqrt{\pi }(-y)^{3}}\Gamma \left( \frac{D+3%
}{2}\right) \Gamma \left( \frac{D}{2}-2\right) ,\;D>4,  \notag \\
G_{D}(y) &\approx &\Gamma (D-1)\frac{\Gamma ((D+3)/2)\Gamma (2-D/2)}{2^{D-3}%
\sqrt{\pi }(-y)^{D-1}},\;D<4,  \label{GDlarge}
\end{eqnarray}%
and%
\begin{equation}
G_{4}(y)\approx 3(-y)^{-3}\left[ 2\ln \left( -4y\right) -3\right] ,\;D=4.
\label{G4large}
\end{equation}%
For the VEV\ of the field squared we get%
\begin{eqnarray}
\langle E^{2}\rangle _{\mathrm{b}} &\approx &\frac{\left( D-1\right) \left(
8-D\right) \Gamma (D/2-2)}{2^{4-D}(4\pi )^{D/2}\alpha ^{D+1}\left(
z^{D}/\eta \right) ^{6}},\;D>4,  \notag \\
\langle E^{2}\rangle _{\mathrm{b}} &\approx &\frac{2^{3-D}\Gamma (D+1)\Gamma
(2-D/2)}{(4\pi )^{D/2}\alpha ^{D+1}\left( z^{D}/\eta \right) ^{2(D-1)}}%
,\;D<4,  \label{E2blarge}
\end{eqnarray}%
and%
\begin{equation}
\langle E^{2}\rangle _{\mathrm{b}}\approx 3\frac{\ln \left( 2z^{D}/\eta
\right) -5/8}{\pi ^{2}\alpha ^{5}\left( z^{D}/\eta \right) ^{6}},\;D=4.
\label{E2bl4}
\end{equation}%
For $D=3$, the expression (\ref{E2blarge}) coincides with the exact result.
In $D=8$ the leading term of (\ref{E2blarge}) vanishes. Numerical
calculations show that for $3\leqslant D\leqslant 8$ the plate-induced part $%
\langle E^{2}\rangle _{\mathrm{b}}$ is positive everywhere. For $D\geqslant
9 $, $\langle E^{2}\rangle _{\mathrm{b}}$ is positive near the plate and
negative at large distances from the plate. In this case, $\langle
E^{2}\rangle _{\mathrm{b}}$ has a minimum for some intermediate value of $%
z^{D}/\eta $.

In a similar way, for the plate-induced part in the VEV of the Lagrangian
density one finds%
\begin{equation}
-\frac{1}{16\pi }\langle F_{\mu \nu }F^{\mu \nu }\rangle _{\mathrm{b}}=\frac{%
1}{8\pi }\left\{ \langle E^{2}\rangle _{\mathrm{b}}+\frac{D-1}{2B_{D}\alpha
^{D+1}}\left[ 2(1-y)\partial _{y}+D-4\right] F_{D}(y)\right\} .  \label{Lb}
\end{equation}%
The second term in the figure braces presents the magnetic contribution to
the Lagrangian density. For $D=3$ it coincides with $\langle E^{2}\rangle _{%
\mathrm{b}}$.

Another important characteristic of the vacuum state is the VEV of the
energy-momentum tensor. In addition to describing the local structure of the
vacuum state, it acts as the source of gravity in the quasiclassical
Einstein equations and plays an important role in modelling self-consistent
dynamics involving the gravitational field. Similar to the case of the field
squared, the VEV of the the energy-momentum tensor is decomposed into the
boundary-free and plate-induced parts:%
\begin{equation}
\langle T_{\mu }^{\nu }\rangle =\langle T_{\mu }^{\nu }\rangle _{0}+\langle
T_{\mu }^{\nu }\rangle _{\mathrm{b}}.  \label{Tdec}
\end{equation}%
Again, for points outside the plate the renormalization is required for the
boundary-free part only. Because of the maximal symmetry of the background
geometry, the latter is proportional to the metric tensor: $\langle T_{\mu
}^{\nu }\rangle _{0}=\mathrm{const}\cdot \delta _{\mu }^{\nu }$.

Here we are interested in the plate-induced part which is directly evaluated
by using the formula
\begin{equation}
\langle T_{\mu }^{\nu }\rangle _{\mathrm{b}}=-\frac{1}{4\pi }\lim_{x^{\prime
}\rightarrow x}\langle F_{\mu }^{\cdot \beta }(x)F_{\cdot \beta }^{\nu
}(x^{\prime })\rangle _{\mathrm{b}}+\frac{\delta _{\mu }^{\nu }}{16\pi }%
\langle F_{\beta \sigma }F^{\beta \sigma }\rangle _{\mathrm{b}}.  \label{Tb}
\end{equation}%
By taking into account the expressions (\ref{F2p}), (\ref{FFb}) and (\ref{Lb}%
), for the VEVs of the diagonal components one finds (no summation over $%
l=1,\ldots ,D-1$)%
\begin{eqnarray}
\langle T_{0}^{0}\rangle _{\mathrm{b}} &=&\frac{\alpha ^{-D-1}}{A_{D}}%
\left\{ \left[ 2\left( 1-y\right) \partial _{y}-D+2\right] G_{D}(y)-\left[
2\left( 1-y\right) \partial _{y}+D-4\right] F_{D}(y)\right\} ,  \notag \\
\langle T_{D}^{D}\rangle _{\mathrm{b}} &=&\frac{\alpha ^{-D-1}}{A_{D}}\left[
2\left( 1-y\right) \partial _{y}-D\right] \left[ F_{D}(y)-G_{D}(y)\right] ,
\notag \\
\langle T_{l}^{l}\rangle _{\mathrm{b}} &=&-\frac{\alpha ^{-D-1}}{A_{D}}\frac{%
D-3}{D-1}\left\{ \left[ 2\left( 1-y\right) \partial _{y}-\left( D-4\right)
\frac{D-1}{D-3}\right] G_{D}(y)\right.  \notag \\
&&\left. +\left[ 2\left( 1-y\right) \partial _{y}+\left( D-4\right) \frac{D-1%
}{D-3}-4\right] F_{D}(y)\right\} ,  \label{Tllb}
\end{eqnarray}%
where%
\begin{equation*}
A_{D}=(4\pi )^{(D+1)/2}(D+1)\Gamma \left( \frac{D-1}{2}\right) .
\end{equation*}%
In addition to the diagonal components, one has also a nonzero off-diagonal
component of the vacuum energy-momentum tensor:%
\begin{equation}
\langle T_{0}^{D}\rangle _{\mathrm{b}}=\frac{\alpha ^{-D-1}}{A_{D}}\frac{%
4z^{D}}{\eta }\left[ (1-y)\partial _{y}-2\right] F_{D}(y).  \label{T0Db}
\end{equation}%
The latter corresponds to the energy flux along the direction normal to the
plate.

As in the case of the field squared, the VEV of the energy-momentum tensor
depends on the coordinates $z^{D}$ and $\eta $ in the combination $%
z^{D}/\eta $. This is a consequence of the maximal symmetry of the
background spacetime and of the Bunch-Davies vacuum state. For $D=3$, by
using the expressions (\ref{F5}) for the functions $F_{D}(y)$ and $G_{D}(y)$%
, we can see that $\langle T_{\mu }^{\nu }\rangle _{\mathrm{b}}=0$. This
result is a direct consequence of the conformal invariance of the
electromagnetic field in $D=3$.

It can be checked that the plate-induced parts of the vacuum energy-momentum
tensor obey the covariant conservation equation $\nabla _{\nu }\langle
T_{\mu }^{\nu }\rangle _{\mathrm{b}}=0$. For the problem under consideration
it reduces to the following two equations:%
\begin{eqnarray}
\eta \partial _{\eta }\langle T_{0}^{0}\rangle _{\mathrm{b}}-D\langle
T_{0}^{0}\rangle _{\mathrm{b}}-\eta \partial _{D}\langle T_{0}^{D}\rangle _{%
\mathrm{b}}+\sum_{k=1}^{D}\langle T_{k}^{k}\rangle _{\mathrm{b}} &=&0,
\notag \\
\eta \partial _{\eta }\langle T_{0}^{D}\rangle _{\mathrm{b}}-\left(
D+1\right) \langle T_{0}^{D}\rangle _{\mathrm{b}}+\eta \partial _{D}\langle
T_{D}^{D}\rangle _{\mathrm{b}} &=&0.  \label{ConsEq}
\end{eqnarray}

Let us consider the behavior of the VEV of the energy-momentum tensor at
small and large distances from the plate. At small distances, $z^{D}/\eta
\ll 1$, by using the asymptotic formulas for the hypergeometric function, to
the leading order one finds (no summation over $l=0,\ldots ,D-1$)%
\begin{equation}
\langle T_{l}^{l}\rangle _{\mathrm{b}}\approx -\frac{\eta }{z^{D}}\langle
T_{0}^{D}\rangle _{\mathrm{b}}\approx \frac{D-1}{(z^{D}/\eta )^{2}}\langle
T_{D}^{D}\rangle _{\mathrm{b}}\approx -\frac{\left( D-3\right) (D-1)\Gamma
((D+1)/2)}{2(4\pi )^{(D+1)/2}\alpha ^{D+1}(z^{D}/\eta )^{D+1}}.
\label{TllbSmall}
\end{equation}%
In this region, all the components are negative for $D\geqslant 4$ and one
has $|\langle T_{0}^{0}\rangle _{\mathrm{b}}|\gg |\langle T_{D}^{D}\rangle _{%
\mathrm{b}}|$. At large distances from the plate, by using the asymptotic
expression%
\begin{equation*}
F_{D}(y)\approx \frac{2^{D-3}}{\sqrt{\pi }y^{2}}\Gamma \left( \frac{D+3}{2}%
\right) \Gamma \left( \frac{D}{2}-1\right) ,\;D\geqslant 3,
\end{equation*}%
valid for $|y|\gg 1$, for $D>4$ one gets (no summation over $l=1,\ldots ,D$)
\begin{eqnarray}
\langle T_{0}^{0}\rangle _{b} &\approx &\frac{D}{D-4}\langle
T_{l}^{l}\rangle _{b}\approx -\frac{2^{D-4}D(D-1)\Gamma (D/2-1)}{(4\pi
)^{D/2+1}\alpha ^{D+1}(z^{D}/\eta )^{4}},  \notag \\
\langle T_{0}^{D}\rangle _{b} &\approx &\frac{2^{D-2}(D-1)\Gamma (D/2-1)}{%
(4\pi )^{D/2+1}\alpha ^{D+1}(z^{D}/\eta )^{5}}.  \label{T0DLarge}
\end{eqnarray}%
For $D=4$ the asymptotic expressions for the components $\langle
T_{0}^{0}\rangle _{\mathrm{b}}$ and $\langle T_{0}^{D}\rangle _{\mathrm{b}}$
are still given by (\ref{T0DLarge}), whereas for the stresses one has (no
summation over $l=1,\ldots ,D$)%
\begin{equation}
\langle T_{l}^{l}\rangle _{\mathrm{b}}=-\frac{3\alpha ^{-5}\ln (z^{D}/\eta )%
}{16\pi ^{3}(z^{D}/\eta )^{6}}.  \label{TllbD4}
\end{equation}%
As it is seen, at large distances the stresses are isotropic.

\begin{figure}[tbph]
\begin{center}
\epsfig{figure=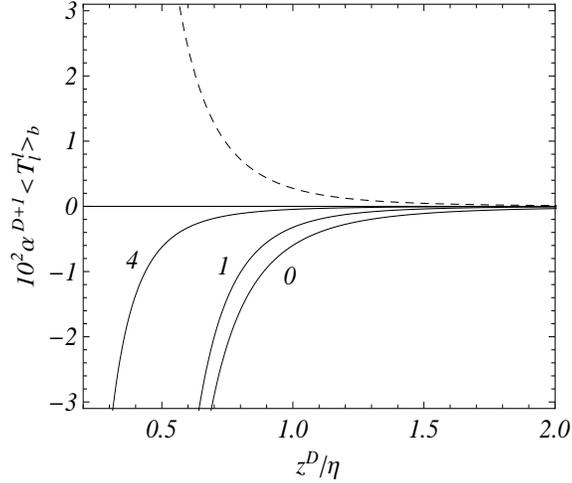,width=7.5cm,height=6.5cm}
\end{center}
\caption{The plate-induced part in the VEV of the energy-momentum tensor as
a function of the proper distance from the plate for $D=4$. The full and
dashed curves correspond to the diagonal and off-diagonal components
respectively. The numbers near the full curves correspond to the values of
the index $l$.}
\label{fig1}
\end{figure}

In figure \ref{fig1} we have plotted the plate-induced parts in the
components of the vacuum energy-momentum tensor as functions of $z^{D}/\eta $
for $D=4$. The full curves correspond to the diagonal components, $\alpha
^{D+1}\langle T_{l}^{l}\rangle _{\mathrm{b}}$, and the numbers near these
curves are the values of the index $l$. The dashed curve corresponds to the
off-diagonal component, $\alpha ^{D+1}\langle T_{0}^{D}\rangle _{\mathrm{b}}$%
. As we see, all the diagonal components of the plate-induced
energy-momentum tensor are negative, whereas the off-diagonal component
corresponding to the energy flux is positive. The numerical calculations
have shown that this is the case for other values of $D$. In particular, the
energy flux is directed from the plate.

Formulas (\ref{Tllb}) and (\ref{T0Db}) present the components of the
energy-momentum tensor in the coordinates $(\tau ,z^{1},\ldots ,z^{D})$. For
the components in the coordinates $(t,z^{1},\ldots ,z^{D})$, denoted as $%
\langle T_{\mathrm{(s)}\mu }^{\nu }\rangle _{\mathrm{b}}$, one has (no
summation over $l=0,1,\ldots ,D$) $\langle T_{\mathrm{(s)}l}^{l}\rangle _{%
\mathrm{b}}=$ $\langle T_{l}^{l}\rangle _{\mathrm{b}}$, $\langle T_{\mathrm{%
(s)}0}^{D}\rangle _{\mathrm{b}}=(\eta /\alpha )\langle T_{0}^{D}\rangle _{%
\mathrm{b}}$ ($\mathrm{s}$ indicates the components in the synchronous time
coordinate). Let $E_{V}^{\text{\textrm{(b)}}}$ be the boundary-induced part
of the vacuum energy (with respect to the time coordinate $t$) in the
spatial volume $V$ with a boundary $\partial V$:%
\begin{equation}
E_{V}^{\text{\textrm{(b)}}}=\int_{V}d^{D}z\sqrt{\gamma }\langle T_{\text{%
\textrm{(s)}}0}^{0}\rangle _{\text{\textrm{b}}},  \label{EbV}
\end{equation}%
where $\gamma $ is the determinant of the spatial metric tensor $\gamma
_{il}=-g_{il}$ and the Latin indices run over $1,2,\ldots ,D$. From the
asymptotic expression (\ref{T0DLarge}) it follows that the plate induced
part in the total energy (per unit surface area of the plate) in the region $%
z_{1}^{D}\leqslant z^{D}<\infty $ is finite. From the equation $\nabla _{\nu
}\langle T_{\text{\textrm{(s)}}\mu }^{\nu }\rangle _{\text{\textrm{b}}}=0$
with $\mu =0$, it follows that%
\begin{equation}
\partial _{t}E_{V}^{\text{\textrm{(b)}}}=-\int_{\partial V}d^{D-1}z\,\sqrt{h}%
n_{l}\langle T_{\text{\textrm{(s)}}0}^{l}\rangle _{\text{\textrm{b}}}+\frac{1%
}{\alpha }\int_{V}d^{D}z\sqrt{\gamma }\langle T_{\text{\textrm{(s)}}%
l}^{l}\rangle _{\text{\textrm{b}}},  \label{EnCons}
\end{equation}%
where $n_{l}$, $\gamma ^{il}n_{i}n_{l}=1$, is the external normal to the
boundary $\partial V$ and $h$ is the determinant of the induced metric $%
h_{il}=\gamma _{il}-n_{i}n_{l}$. The first term in the right-hand side of
Eq. (\ref{EnCons}) describes the energy flux through the boundary $\partial
V $. As a volume $V$ let us take a cylinder with the axis perpendicular to
the plate and with the bases at $z_{1}^{D}$ and $z_{2}^{D}$. If $S$ is the
area of the cylinder base, $S=\int dz^{1}\cdots dz^{D-1}$, then the proper
area is given by $(\alpha /\eta )^{D-1}S$. With this choice of the volume $V$
we have $n_{l}|_{z^{D}=z_{j}^{D}}=(-1)^{j}\delta _{l}^{D}\alpha /\eta $, and%
\begin{equation*}
\int_{\partial V}d^{D-1}z\,\sqrt{h}n_{l}\langle T_{\text{\textrm{(s)}}%
0}^{l}\rangle _{\text{\textrm{b}}}=(\alpha /\eta )^{D}S\left[ \langle T_{%
\text{\textrm{(s)}}0}^{D}\rangle _{\text{\textrm{b}}}|_{z^{D}=z_{2}^{D}}-%
\langle T_{\text{\textrm{(s)}}0}^{D}\rangle _{\text{\textrm{b}}%
}|_{z^{D}=z_{1}^{D}}\right] .
\end{equation*}%
From here it follows that $\langle T_{0}^{D}\rangle _{\text{\textrm{b}}%
}=(\alpha /\eta )\langle T_{\text{\textrm{(s)}}0}^{D}\rangle _{\text{\textrm{%
b}}}$ is the energy flux per unit proper surface area. In particular, for
the energy in the region $z^{D}\geqslant z_{1}^{D}$ we get%
\begin{equation}
\partial _{t}E_{z^{D}\geqslant z_{1}^{D}}^{\text{\textrm{(b)}}}=(\alpha
/\eta )^{D-1}S\Big[\langle T_{0}^{D}\rangle _{\text{\textrm{b}}%
}|_{z^{D}=z_{1}^{D}}+\frac{1}{\eta }\int_{z_{1}^{D}}^{\infty }dz^{D}\langle
T_{l}^{l}\rangle _{\text{\textrm{b}}}\Big].  \label{dtEinf}
\end{equation}%
The first term in the square brackets of (\ref{dtEinf}) is positive whereas
the second one is negative.

\section{Conclusion}

\label{sec:Conc}

In the present paper we have considered the two-point functions for the
electromagnetic field in background of dS spacetime assuming that the field
is prepared in the Bunch-Davies vacuum state. To this aim, a complete set of
mode functions was constructed. Then we have evaluated the two-point
functions in the geometry of a conducting plate. By using these functions
the VEVs of the field squared and the energy-momentum tensor are
investigated. These VEVs are decomposed into the boundary-free and
plate-induced parts. For points outside of the plate the renormalization is
needed for the first parts only. Because of the maximal symmetry of the
background spacetime and of the Bunch-Davies vacuum state the boundary-free
parts do not depend on spacetime coordinates. The plate-induced parts depend
on the coordinates $z^{D}$ and $\eta $ in the form of the ratio $z^{D}/\eta $%
. The latter is the proper distance of the observation point from the plate,
measured in the units of the dS curvature scale $\alpha $. The plate-induced
part in the VEV of the electric field squared is given by formula (\ref{E2b}%
). For $3\leqslant D\leqslant 8$ this contribution is positive everywhere,
whereas for $D\geqslant 9$ it is positive near the plate and negative at
large distances. Simple asymptotic expressions, (\ref{E2bAs}), (\ref%
{E2blarge}), (\ref{E2bl4}), are obtained at small and at large distances
from the plate. At large distances, the plate-induced part decays as $%
(z^{D}/\eta )^{-6}$ for $D>4$ and as $(z^{D}/\eta )^{2(1-D)}$ for $D<4$. For
$D=4$ one has the asymptotic behavior $(z^{D}/\eta )^{-6}\ln (2z^{D}/\eta )$.

The plate-induced parts in the VEVs of the diagonal components of the
energy-momentum tensor are given by the expressions (\ref{Tllb}). In
addition to these components we have also nonzero off-diagonal component (%
\ref{T0Db}) which describes energy flux along the direction normal to the
plate. The plate-induced part in the VEV of the energy-momentum tensor
vanishes for $D=3$. In this case the electromagnetic field is conformally
invariant and this result is directly obtained from the corresponding result
for a perfectly conducting plate in Minkowski spacetime by using the
standard conformal transformation. For the components of the vacuum
energy-momentum tensor one has asymptotic expressions (\ref{TllbSmall}), (%
\ref{T0DLarge}) and (\ref{TllbD4}). Near the plate the stresses along the
directions parallel to the plate are equal to the energy density, whereas
for the normal stress and the energy flux one has $\langle T_{D}^{D}\rangle
_{b}\approx (z^{D}/\eta )^{2}\langle T_{0}^{0}\rangle _{b}/(D-1)$ and $%
\langle T_{0}^{D}\rangle _{b}\approx -(z^{D}/\eta )\langle T_{0}^{0}\rangle
_{b}$. At large distances from the plate the vacuum stresses are isotropic.
For $D>4$ the diagonal components of the plate-induced parts decay as $%
\left( z^{D}/\eta \right) ^{-4}$ and the off-diagonal component decays like $%
(z^{D}/\eta )^{-5}$. For $D=4$ the asymptotic behavior for the components $%
\langle T_{0}^{0}\rangle _{b}$ and $\langle T_{0}^{D}\rangle _{b} $ remain
the same, whereas the stresses behave as $(z^{D}/\eta )^{-6}\ln \left(
z^{D}/\eta \right) $. For the numerical examples we have considered, all the
diagonal components of the plate-induced vacuum energy-momentum tensor are
negative, whereas the off-diagonal component $\langle T_{0}^{D}\rangle _{b}$
is positive. In particular, the energy flux is directed from the plate.

We have considered here the electromagnetic field interacting with
boundaries and with the background gravitational field only. An interesting
development of the problem in question would be the investigation of the
role of loop corrections on the Casimir effect in dS spacetime. For a
self-interacting scalar field the loop corrections on boundary-free dS
spacetime have been attracting a lot of attention recently (see, for
example, the recent review \cite{Akhm13}). The calculations have shown that
loop diagrams typically exhibit large infrared logarithms which are of key
importance in discussing quantum field theory on dS background.

\section*{Acknowledgments}

This work was supported by State Committee Science MES RA, within the frame
of the research project No. SCS 13-1C040. A.A.S. gratefully acknowledges the
hospitality of the INFN Laboratori Nazionali di Frascati (Frascati, Italy)
where part of this work was done.

\end{document}